\documentclass[aps,showpacs,nofootinbib]{revtex4}
\usepackage{bm}
\usepackage{graphics}
\usepackage{epsfig}
\usepackage{graphicx}
\usepackage{amsmath}
\usepackage{amsfonts}
\usepackage{amssymb}
\usepackage{lineno}
\begin{document}

\title{Spin-glass instability of short-range spherical ferromagnet}
\author{P. N. Timonin}
\email{timonin@aaanet.ru}
\affiliation{Physics Research Institute at
Southern Federal University, 344090 Rostov-on-Don, Russia}

\date{\today}

\begin{abstract}
In structurally disordered ferromagnets the weak random dipole-dipole exchange may transform the polydomain state into a spin-glass one. To some extent the properties of such phase in disordered isotropic ferromagnet can be qualitatively described by the spherical model with the short-range ferromagnetic interaction and weak frustrated infinite-range random-bond exchange. This model is shown to predict that spin-glass phase substitute the ferromagnetic one at the arbitrary small disorder strength and that its thermodynamics has some similarity to that of polydomain state along with some significant distinctions. In particular, the longitudinal susceptibility at small fields becomes frozen below transition point at a constant value depending on the disorder strength, while the third order nonlinear magnetic susceptibilitiy exhibits the temperature oscillations in small field near the transition point. The relation of these predictions to the experimental data for some disordered isotropic ferromagnets is discussed. 
\end{abstract}

\pacs{64.60.Cn, 05.70.Jk, 64.60.Fr }

\maketitle

The spherical model with short-range exchange shares the basic qualitative features with real isotropic ferromagnets. It has phase transition only in space dimensions $d > 2$ and its coercive field is strictly zero. This is because it is equivalent to the isotropic $n$-component model in the limit $ n\to \infty $ \cite{1}. So the scalar magnetization of short-range spherical model corresponds to the magnetization module of isotropic ferromagnets and this makes this model very useful for the studies of qualitative features of their thermodynamics.
Yet real ferromagnets have also the long-range dipole-dipole interaction. Being a weak relativistic effect it nevertheless determines crucially the nature of ferromagnetic transition which usually results in the appearance of inhomogeneous polydomain state. It shows up in the freezing of longitudinal magnetic susceptibility at the value   $\chi  = (4\pi \kappa)^{-1}$ below $T_c$ at fields   $H < 4\pi \kappa M_s $, $\kappa$ is the depolarizing coefficient along the field direction, $M_s $ is the spontaneous magnetization \cite{2,3}. It is rather natural to suppose that when some non-magnetic disorder such as structural defects or non-magnetic impurities is present in a crystal the polydomain state may transform into the spin-glass one \cite{4}. 

To describe the qualitative features of such spin-glass state in random isotropic ferromagnets we may turn to the spherical model with weak long-range frustrated disorder imitating the random dipole-dipole exchange in the structurally disordered media. The influence of such (infinite-range) disorder on the thermodynamics of the mean-field spherical ferromagnet was studied in Ref. \onlinecite{5}. In this model the spin-glass phase instead of ferromagnetic one do appear when disorder becomes sufficiently strong. Here we consider more realistic short-range spherical model of ferromagnet with the same infinite-range frustrated random exchange. We find that contrary to the mean-field model in the short-range one the spin-glass substitutes the ferromagnetic phase at arbitrary weak random exchange. We also show that the magnetic properties of this spin-glass phase in small magnetic fields have some similarity to those of polydomain ferromagnetic state along with some significant distinctions.
 
The Hamiltonian of the spherical model has the form
\[
{\cal H} =  - \frac{1}{2}{\bf S}\hat J{\bf S} - {\bf HS}.
\]
Here $\bf S $ is $N$-component vector subjected to the constraint ${\bf S}^2  = N$, $\bf H $ is the external field and $J_{i,j}$ is the matrix of exchange integrals. 

Partition sum of the model can be represented as
\begin{eqnarray}
Z = \int\limits_{a - i\infty }^{a + i\infty } {\frac{{d\lambda }}{{2\pi i}}} \exp \left[ { - N\beta F(\lambda )} \right],
\label{eq:1}
\\
 - 2\beta F(\lambda ) = \lambda  - N^{ - 1} Tr\ln \hat G^{ - 1} \left( \lambda  \right) + N^{ - 1} \beta ^2 {\bf H}\hat G\left( \lambda  \right){\bf H},\label{eq:2}
\\
\hat G\left( \lambda  \right) = \left( {\lambda \hat I - \beta \hat J} \right)^{ - 1}\label{eq:3}. 
\end{eqnarray}
 $\beta  = 1/T$ is the inverse temperature.  The parameter $a$ in the integral over $\lambda$ can be arbitrary provided it obeys the condition  $a > \beta J_{\max } $, $J_{\max }$ being the largest eigenvalue of $\hat J$. Thus Eqs. (\ref{eq:1}-\ref{eq:3}) are valid for any $\hat J$ with the spectrum limited from above.  
For the equilibrium thermodynamic potential $F $ we have from Eq.(\ref{eq:1}) at $N \to \infty $
\begin{equation}
F = \mathop {min}\limits_\lambda  F\left( \lambda  \right) = F\left[ {\lambda _0 \left( {\hat J} \right)} \right]. \label{eq:4}
\end{equation}	
Here   $\lambda _0 \left( {\hat J} \right)$ is the value which provides the minimum of $F\left( \lambda  \right)$. It obeys the equation of state
\begin{equation}
\frac{{\partial F\left( {\lambda _0 } \right)}}{{\partial \lambda _0 }} = 0\label{eq:5}
\end{equation}	
Solving Eq. (\ref{eq:5}) and substituting the   $\lambda _0 \left( {\hat J} \right)$ found into Eq. (\ref{eq:3}) we get the equilibrium potential $F$ and can then find all thermodynamic variables of the system. In particular, we get for the average local spins
\begin{equation}
\left\langle {\bf S} \right\rangle _T  =  - N\frac{{\partial F}}{{\partial {\bf H}}} = \beta \hat G\left( {\lambda _0 } \right){\bf H}\label{eq:6}
\end{equation}	
When   $\hat J$ is a random matrix we should average $F$ over it. It can be easily done if we assume  $\lambda _0 \left( {\hat J} \right)$  to be the self-averaging quantity. Then while averaging of Eqs. (4, 5) we can just substitute  $\lambda _0 \left( {\hat J} \right)$  by its average value   $\bar \lambda  = \left\langle {\lambda _0 \left( {\hat J} \right)} \right\rangle _J $. Thus we get from Eqs. (\ref{eq:2}, \ref{eq:4})
\begin{equation}
 - 2\beta \bar F \equiv  - 2\beta \left\langle {F(\lambda )} \right\rangle _J  = \bar \lambda  - \int {d\varepsilon \rho \left( \varepsilon  \right)} \ln (\bar \lambda  - \varepsilon ) + N^{ - 1} \beta ^2 {\bf H}\left\langle {\hat G\left( {\bar \lambda } \right)} \right\rangle _J {\bf H},\label{eq:7}
\end{equation}	

where $\rho \left( \varepsilon  \right)$ is the average spectral density of the matrix  $\beta \hat J$,
\begin{equation}
\rho \left( \varepsilon  \right) = \frac{1}{{\pi N}}\mathop {\lim }\limits_{\delta  \to 0} {\mathop{\rm Im}\nolimits} Tr\left\langle {\hat G\left( {\varepsilon  - i\delta } \right)} \right\rangle _J. \label{eq:8}
\end{equation}	 

From Eqs. (\ref{eq:2}, \ref{eq:3}, \ref{eq:5}) we get the equation for $\bar \lambda $,
\begin{eqnarray}
D\left( {\bar \lambda } \right) + Q\left( {\bar \lambda } \right) = 1,\label{eq:9}
\\
D\left( {\bar \lambda } \right) \equiv N^{ - 1} Tr\left\langle {\hat G\left( {\bar \lambda } \right)} \right\rangle _J,\label{eq:10} 
\\
Q\left( {\bar \lambda } \right) \equiv N^{ - 1} \beta ^2 {\bf H}\left\langle {\hat G^2 \left( {\bar \lambda } \right)} \right\rangle _J {\bf H} = N^{ - 1} \left\langle {\left\langle {\bf S} \right\rangle _T^2 } \right\rangle _J \label{eq:11}
\end{eqnarray}
The last equality in Eq. (\ref{eq:11}) follows from Eq. (\ref{eq:6}). It shows that   $Q\left( {\bar \lambda } \right)$ is the Edwards-Anderson spin-glass order parameter.

	Here we consider the Gaussian disorder for the exchange integrals with the mean
\[
\left\langle {J_{i,j} } \right\rangle  = \bar J\left( {{\bf r}_i  - {\bf r}_j } \right)
\]
and the deviation
\[
\left\langle {\left( {J'_{i,j} } \right)^2 } \right\rangle  = \frac{{\Delta ^2 }}{N},
\qquad
{\rm     }J'_{i,j}  \equiv J_{i,j} {\rm  -  }\bar J\left( {{\bf r}_i  - {\bf r}_j } \right)
\]
We assume $\bar J\left( {{\bf r}_i  - {\bf r}_j } \right)$ to describe the short range ferromagnetic interactions so its Fourier transform $\bar J\left( {\bf k} \right)$ have a maximum at k = 0 and near it
\[
\bar J\left( {\bf k} \right) \approx \bar J - Ak^2. 
\] 
Then on a three-dimensional lattice the spectral density of  $ \beta \bar J\left( {\bf k} \right)$,
\[
\rho _0 \left( \varepsilon  \right) = \int {\frac{{d^3 k}}{{\left( {2\pi } \right)^3 }}} \delta \left[ {\varepsilon  - \beta \bar J\left( {\bf k} \right)} \right],
\]
would have the square-root behavior at the upper edge of the spectrum which describes the most relevant long-range ferromagnetic fluctuations,
\[
\rho_0 \left( \varepsilon  \right) \sim \sqrt {\beta \bar J - \varepsilon }. 
\]
So we choose 
\begin{equation}
\rho_0 \left( \varepsilon  \right) = \frac{2}{{\pi \left( {\beta \bar J} \right)^2 }}\vartheta \left[ {\left( {\beta \bar J} \right)^2  - \varepsilon ^2 } \right]\sqrt {\left( {\beta \bar J} \right)^2  - \varepsilon ^2 }. 
\label{eq:12}
\end{equation}
Here $\theta$ is the Heaviside step function. This $\rho _0 \left( \varepsilon  \right)$ correctly behaves at the upper edge and makes further calculations quite easy. The explicit form of $\bar J\left( {\bf k} \right)$ appears to be irrelevant for the homogeneous external field we consider below and all thermodynamics is determined solely by  $\rho _0 \left( \varepsilon  \right)$.

	Now we can find $\left\langle {\hat G\left( {\bar \lambda } \right)} \right\rangle _J$ for such random ensemble where the weak infinite-range random exchange fluctuations of arbitrary sign coexist with non-random short-range ferromagnetic interactions. Expanding $\hat G\left( {\bar \lambda } \right)$ in the power series of $J'_{i,j}$ and averaging this expansion with the Gaussian distribution we find in the large $N$ limit the following expression for the Fourier transform of $\left\langle {\hat G\left( {\bar \lambda } \right)} \right\rangle _J$,
\begin{equation}
\bar G^{ - 1} \left( {\bar \lambda ,{\bf k}} \right) = \bar \lambda  - \beta ^2 \Delta ^2 D\left( {\bar \lambda } \right) - \beta \bar J\left( {\bf k} \right).
\label{eq:13}
\end{equation}
Then for $D\left( {\bar \lambda } \right)$ (\ref{eq:10}) we have the equation
\[
D\left( \bar \lambda \right) = \int {d\varepsilon \frac{\rho_0 \left( \varepsilon  \right)}{\bar \lambda  - \beta ^2 \Delta ^2 D\left( \bar \lambda  \right) - \varepsilon }}= 
 \frac{2}{\left( \beta \bar J \right)^2 }\left[ \bar \lambda  - \beta ^2 \Delta ^2 D\left( \bar \lambda  \right) - \sqrt {\left[ \bar \lambda  - \beta ^2 \Delta ^2 D\left( \bar \lambda \right) \right]^2  - \left( \beta \bar J \right)^2} \right]
\]
The solution to this equation is
\begin{eqnarray}
D\left( {\bar \lambda } \right) = \frac{{2c^2 }}{{\left( {\beta \bar J} \right)^2 }}\left[ {\bar \lambda  - \sqrt {\bar \lambda ^2  - c^{ - 2} \left( {\beta \bar J} \right)^2 } } \right], 
\qquad
c^2  \equiv \left( {1 + 4\frac{{\Delta ^2 }}{{\bar J^2 }}} \right)^{ - 1}
\label{eq:14}
\end{eqnarray}
Eqs. (\ref{eq:13}, \ref{eq:14}) define  $\bar G\left( {\bar \lambda ,{\bf k}} \right)$. From these equations we can also find the Fourier transform of  $\left\langle {\hat G^2 \left( {\bar \lambda } \right)} \right\rangle _J $,
\begin{equation}
\bar G_2 \left( {\bar \lambda ,{\bf k}} \right) =  - \frac{\partial }{{\partial \bar \lambda }}\bar G\left( {\bar \lambda ,{\bf k}} \right) = \bar G^2 \left( {\bar \lambda ,{\bf k}} \right)\left[ {1 - \beta ^2 \Delta ^2 D'\left( {\bar \lambda } \right)} \right].
\label{eq:15}
\end{equation}
From (\ref{eq:8}, \ref{eq:10}, \ref{eq:14}) we also get
\begin{equation}
\rho \left( \varepsilon  \right) = \frac{1}{\pi }\mathop {\lim }\limits_{\delta  \to 0} {\mathop{\rm Im}\nolimits} D\left( {\varepsilon  - i\delta } \right) = 
 \frac{{2c^2 }}{{\pi \left( {\beta \bar J} \right)^2 }}\vartheta \left[ {c^{ - 2} \left( {\beta \bar J} \right)^2  - \varepsilon ^2 } \right]\sqrt {c^{ - 2} \left( {\beta \bar J} \right)^2  - \varepsilon ^2 } \label{eq:16}
\end{equation} 
Thus we have all that is needed to obtain the explicit expressions for the average thermodynamic potential (\ref{eq:7}) and the equation of state (\ref{eq:9}). Further we consider the homogeneous external field, 
$H_i = H, i = 1,…, N$. It is convenient to introduce the new variable $z, 0 < z < 1$, instead of  $\bar \lambda$,
\begin{equation}
\bar \lambda  = \frac{{\beta \bar J}}{{2c}}\left( {z^{ - 1}  + z} \right)
\label{eq:17}
\end{equation}
Then we have from Eqs. (9-11, 13-15, 17) the equation which defines $z$,
\begin{eqnarray}
h^2 z^2 \left( {1 + cz} \right) = \left( {1 - tz} \right)\left( {1 - z^2 } \right)\left( {1 - cz} \right)^3, \label{eq:18} 
\\
h \equiv H/T_g ,{\rm    }T_g  \equiv \sqrt {\bar J^2 /4 + \Delta ^2 } ,{\rm      }t \equiv T/T_g. \nonumber
\end{eqnarray}

From Eqs. (\ref{eq:7}, \ref{eq:14}, \ref{eq:16}, \ref{eq:17}) we get the average potential,
\begin{equation}
 - 2\bar F/T_g  = t\ln t + z + z^{ - 1}  + \frac{{zh^2 }}{{\left( {1 - cz} \right)^2 }} + t\left( {\ln z - \frac{{z^2 }}{2}} \right)
 \label{eq:19}
\end{equation}
It can be easily checked that Eq. (\ref{eq:18}) is equivalent to the equation $\frac{{\partial \bar F}}{{\partial z}} = 0$ and that the solution of it provides the minimum of potential in the interval $0 < z < 1$.
Other thermodynamic parameters can be also expressed via $z$. Thus averaging Eq. (\ref{eq:6}) over random exchange we get the average magnetization
\begin{equation}
M = \beta H\bar G\left( {\bar \lambda ,{\bf k} = 0} \right) = \frac{{zh}}{{\left( {1 - cz} \right)^2 }}
\label{eq:20},
\end{equation}	
while from (\ref{eq:9}, \ref{eq:10}, \ref{eq:17}) we get for the equilibrium value of the Edwards-Anderson order parameter
\begin{equation}
Q = 1 - tz.
\label{eq:21}
\end{equation}		
Also from Eq. (\ref{eq:19}) we obtain the entropy
\begin{equation}
S = \frac{1}{2}\left( {1 + \ln tz - \frac{{z^2 }}{2}} \right)
\label{eq:22},
\end{equation}
and the heat capacity
\begin{equation}
C = \frac{1}{2}\left[ {1 + \left( {1 - z^2 } \right)\frac{{d\ln z}}{{d\ln t}}} \right].
\label{eq:23}
\end{equation}	
Thus Eqs. (\ref{eq:20}-\ref{eq:23}) supplied with the solution to Eq. (\ref{eq:18}) for   $z = z\left( {t,h,c} \right)$ give full description of the thermodynamics of the model. Here we should note that parameter $c$ defined in Eq. (\ref{eq:14}) determine the relative strength of the short-range ferromagnetic bonds. It varies in the interval   $0 \le c \le 1$ and $c = 1$ corresponds to the pure short-range ferromagnet while at $c = 0$ only random infinite-range glassy exchange is present in the system. So at $c = 1$ we have ordinary ferromagnetic transition at $t = 1, h =0$ with anomalies usual to the pure spherical model. In this case $Q = M^2$. Yet at all $c < 1$ this transition is destroyed and instead the transition into the spin-glass phase takes place at $t = 1, h = 0$.

Indeed, when $h \to 0$ then $z \to 1/t$ for $t  > 1$ and $z \to 1$ for $t  < 1$. So at all $c < 1$ and $h  = 0$ $M$ is zero, but spontaneous $Q$ appears at $t  < 1$, $Q = 1- t $. This is in sharp contrast with the model where instead of short-range $\bar J\left( {\bf k} \right)$ the infinite-range mean-field ferromagnetic interaction of the form $\bar J\left( {\bf k} \right) = \delta _{{\bf k},0} \bar J/N$ is introduced \cite{5}. Then spin-glass transition substitutes the ferromagnetic one only at $c^2 < 1/5$.
  
As the short-range spherical ferromagnet correctly reproduces the qualitative features of real isotropic ferromagnets we may suppose tentatively that the destruction of magnetic order by the infinitesimal glassy long-range random-bond disorder can also take place in real isotropic magnets with structural imperfections. In such a case the present model can reveal the qualitative features accompanying this phenomenon in some amorphous ferromagnets, ferromagnetic alloys and even in the nominally pure ferromagnetic crystals.
 
\begin{figure}[htp]
\centering
\includegraphics[height=6cm]{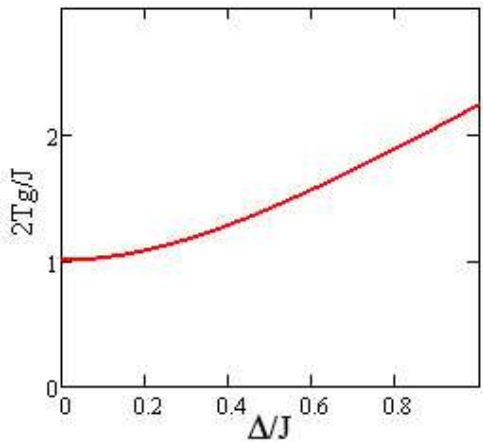}
\caption{(color online) The dependence of $T_g$ on the relative disorder strength $\Delta /\bar J$ .}\label{Fig.1}
\end{figure}

The dependence of the spin-glass transition temperature $T_g$ (18) on the relative strength ($\Delta /\bar J$) of frustrated disorder is shown in Fig. \ref{Fig.1}.

At $h = 0$ we have for all $t$
\begin{eqnarray*}
S = \left\{ \begin{array}{l}
 \frac{1}{2}\left( {1 - \frac{1}{{2t^2 }}} \right),{\rm   }t > 1 \\ 
 \frac{1}{2}\left( {\frac{1}{2} + \ln t} \right),{\rm   }t < 1 \\ 
 \end{array} \right.
\qquad
C = \left\{ \begin{array}{l}
 \frac{1}{{2t^2 }},{\rm   }t > 1 \\ 
 \frac{1}{2},{\rm   }t < 1 \\ 
 \end{array} \right.
\qquad
\chi  \equiv \frac{{\partial M}}{{\partial h}} = \left\{ \begin{array}{l}
 \frac{t}{{\left( {t - c} \right)^2 }},{\rm   }t > 1 \\ 
 \frac{1}{{\left( {1 - c} \right)^2 }},{\rm   }t < 1 \\ 
 \end{array} \right.
 \end{eqnarray*} 
      
\begin{figure}[htp]
\centering
\includegraphics[height=6cm]{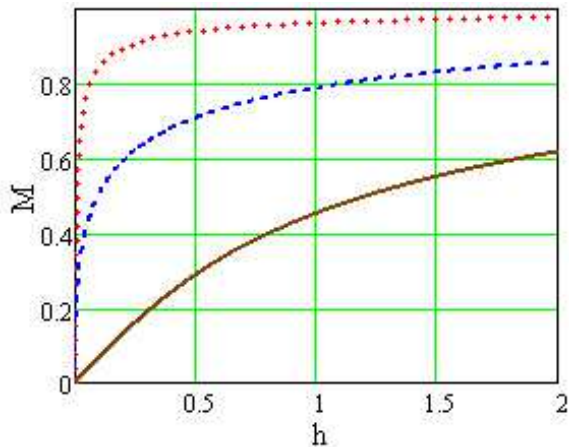}
\caption{(color online) Field dependence of magnetization for $c = 0.9$ at various temperatures, $t = 3$ (solid line), $t=1$ (dashed line), $t=0.1$ (dotted line).}\label{Fig.2}
\end{figure}

\begin{figure}[htp]
\centering
\includegraphics[height=6cm]{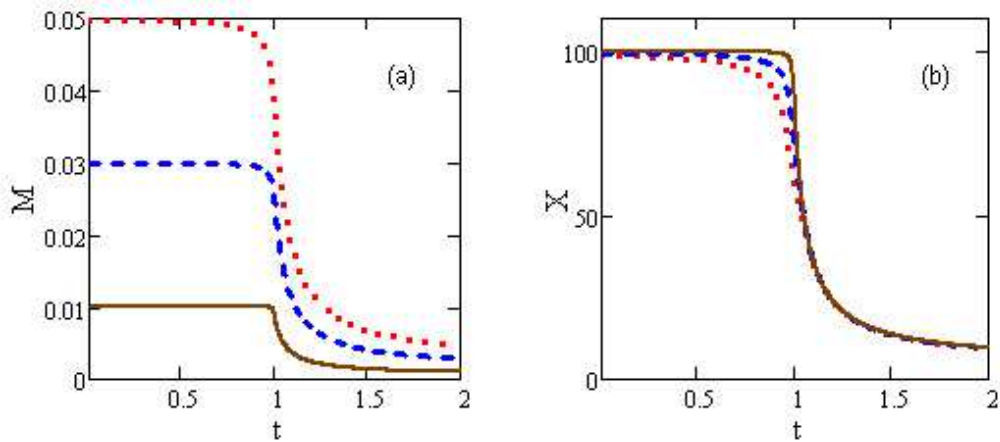}
\caption{(color online) Temperature dependencies of $M $ (a) and $\chi$ (b) for $c = 0.9$ in small fields, $h =5\times 10^{-4}$ (dotted lines), $3\times 10^{-4}$ (dashed lines), $10^{-4}$ (solid lines).}\label{Fig.3}
\end{figure}

So the zero-field thermal properties of the model do not depend on $c$, while the magnetic susceptibility $\chi$ is essentially defined by it. Fig. \ref{Fig.2} shows the field dependence of magnetization at various temperatures. Note the steep rise of $M$ at low fields below $T_g$. Here the slope of $M(h)$ in small fields is limited by the value of zero-field susceptibility, $(1-c)^{-2}$  , while in the polydomain ferromagnet it is limited by   $\chi = (4\pi\kappa)^{-1}$. 
We can easily find $M(t, h)$ from  Eqs. (\ref{eq:18}, \ref{eq:20}) for small fields  $h^2  \ll t\left( {1 - c} \right)^3 $,
\begin{equation}
M = h\left[ {\left( {1 - c} \right)^2  + \left( {1 - c^2 } \right)\left( {\tau  + \sqrt {\tau ^2  + bh^2 } } \right)} \right]^{ - 1} 
\qquad
\tau  \equiv \frac{{t - 1}}{{2t}},{\rm    }
\qquad
b \equiv \frac{{1 + c}}{{2t\left( {1 - c} \right)^3 }}.
\label{eq:24}
\end{equation}	
Fig. \ref{Fig.3} presents this $M(t)$ and $\chi (t)$  in small fields for $c = 0.9$. They are rather similar to those of pure ferromagnet undergoing the transition into polydomain state albeit with the disorder-dependent saturation values.

\begin{figure}[htp]
\centering
\includegraphics[height=6cm]{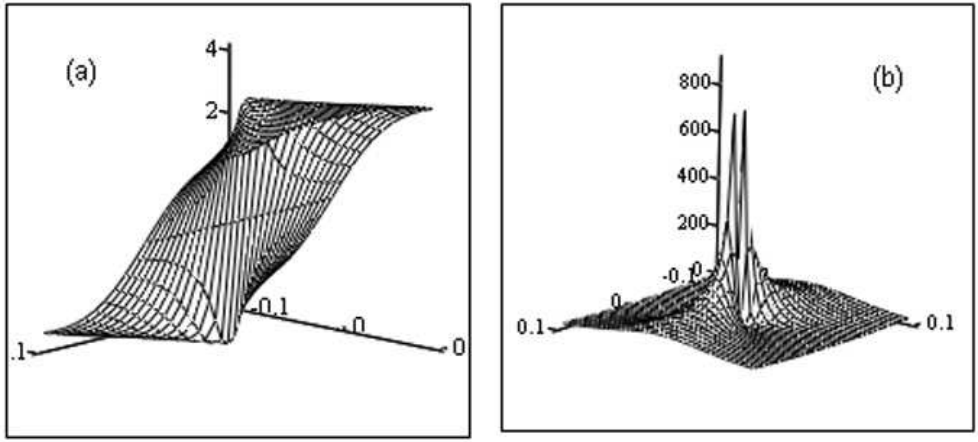}
\caption{(color online) Nonlinear susceptibilities $\chi_2/b^{3/2}$ (a) and $\chi_3/b^2$ (b) near the singular point $\tau = 0, h = 0$.}\label{Fig.4}
\end{figure}

Yet more spectacular anomalies are exhibited by the nonlinear magnetic susceptibilities of the model. They are known to diverge at spin-glass transition in zero field in various mean-field spin-glass models \cite{6} including the spherical one \cite{7}. In the last case these divergences result from the specific non-analyticity of $M( t, h, c)$ at  $t  = 1, h = 0$ in Eq. (\ref{eq:24}) which also give rise to temperature and field oscillations of nonlinear susceptibilities near the transition. Near transition point at $c \ne 1$ and for  $\left| \tau  \right| \ll 1 - c$ we get from Eq. (\ref{eq:24}) two first nonlinear magnetic susceptibilities,
\begin{equation}
\chi _2  \equiv  - \frac{{\partial ^2 M}}{{\partial h^2 }} = \frac{{2b^2 h\left( {3\tau ^2  + 2bh^2 } \right)}}{{\left( {\tau ^2  + bh^2 } \right)^{3/2} }}
\qquad
\chi _3  \equiv  - \frac{{\partial ^3 M}}{{\partial h^3 }} = \frac{{6b^2 \tau ^4 }}{{\left( {\tau ^2  + bh^2 } \right)^{5/2} }}
\label{eq:25}
\end{equation}	
They exhibit highly anisotropic behavior near the singular point $\tau = 0$, $h = 0$. In the polar coordinates defined as
\[
r = \sqrt {\tau ^2  + bh^2 } 
\qquad
\varphi  = \tan ^{ - 1} \left( {\frac{{\sqrt b h}}{\tau }} \right)
\]
we have
\begin{equation}
\chi _2  = 2b^{3/2} \sin \varphi \left( {2 + \cos ^2 \varphi } \right),{\rm   }\chi _3  = \frac{{6b^2 }}{r}\cos ^4 \varphi 
\label{eq:26}
\end{equation}	
Thus at $\varphi = 0 (h = 0)$   $\chi_2=0, \chi_3 =6b^2 \left| \tau  \right|^{-1}$ , while at $\varphi = \pi/2 (\tau = 0) $ $\chi_2=2b^{3/2}sign(h)$, $\chi_3=0$.  The behavior of $\chi_2$ and $\chi_3$ near the singular point $\tau = 0, h = 0$ is shown in Fig. \ref{Fig.4}.
                                                                         
\begin{figure}[htp]
\centering
\includegraphics[height=7cm]{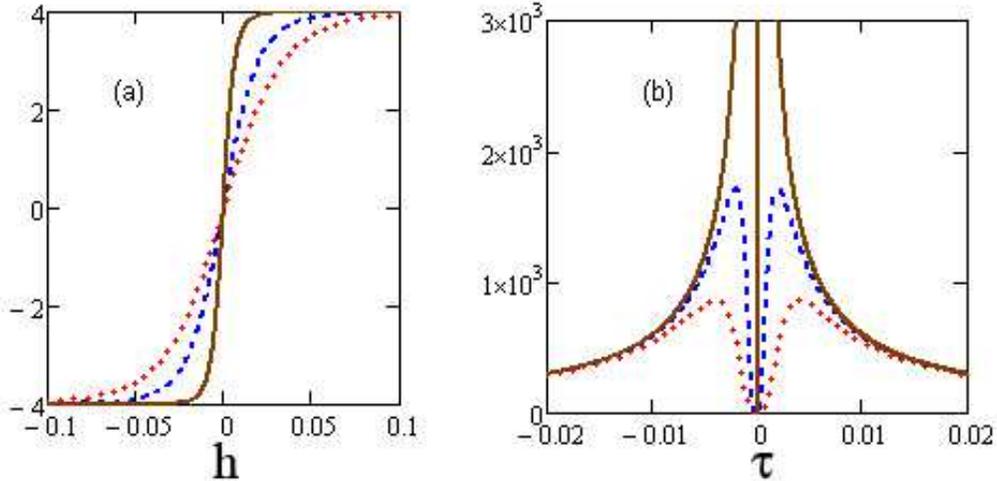}
\caption{(color online)(a) - field dependencies of  $\chi_2/b^{3/2}$    at $\tau$ = 0.01 (solid line), 0.03 (dashed line), 0.05 (dotted line); (b) - temperature dependencies of  $\chi_3/b^2$     at small fields $h$ = 0 (solid line), 0.001 (dashed line), 0.002 (dotted line).}\label{Fig.5}
\end{figure}
	These complex anomalies result in specific field dependence of $\chi_2 $ and temperature oscillations of $\chi_3$ as seen in Fig. \ref{Fig.5}.
	 
	The behavior of nonlinear susceptibilities similar to that of Fig. \ref{Fig.5} is observed in isotropic ferromagnet $Nd_{0.75} Ba_{0.25} MnO_3 $ \cite{8} and in the polycrystalline samples of $RuSr_2GdCu_2O_8$\cite{9}. In the toroidal polycrystalline samples of $La_{0.66} Ba_{0.34} MnO_3 $ with demagnetization factor $\kappa  \approx 0$ the plateau in $\chi(T)$ same as in Fig. \ref{Fig.3}(b) is found manifesting the transition into the glass state \cite{10}. There are many other examples of such step-like behavior of $\chi(T)$ in disordered isotropic magnets, see, for example, Refs. \cite{11}, \cite{12}. But it is often impossible to check the relation $\chi=(4\pi\kappa)^{-1}$ below $T_c$ to distinguish between the polydomain and the spin-glass states as some experimental papers lacks the values of $\kappa$ calculated from the sample shape. It is quite possible that such check will show that many allegedly polydomain ferromagnets are actually the spin-glasses.
	
	Yet now it is not clear if the present result on the spin-glass instability of spherical magnet does apply to the real dipolar Heisenberg magnets which may have some threshold disorder strength to become the spin-glasses. To resolve this issue further theoretical studies of the role of random dipole-dipole interaction in isotropic ferromagnets are needed.
	
	I gratefully acknowledge the useful discussions with V.B. Shirokov and E.D. Gutlianskii.


%
%
\end{document}